\documentclass[twocolumn]{aastex631}

\usepackage{amsmath}
\usepackage{ulem}
\usepackage{xcolor}   
\usepackage{url}
\usepackage{soul}

\catcode`_=\active
\newcommand_[1]{\ensuremath{\sb{\mathrm{#1}}}}
\catcode`^=\active
\newcommand^[1]{\ensuremath{\sp{\mathrm{#1}}}}

\newcommand{\figone}
{\begin{figure*}
\includegraphics[width=\linewidth]{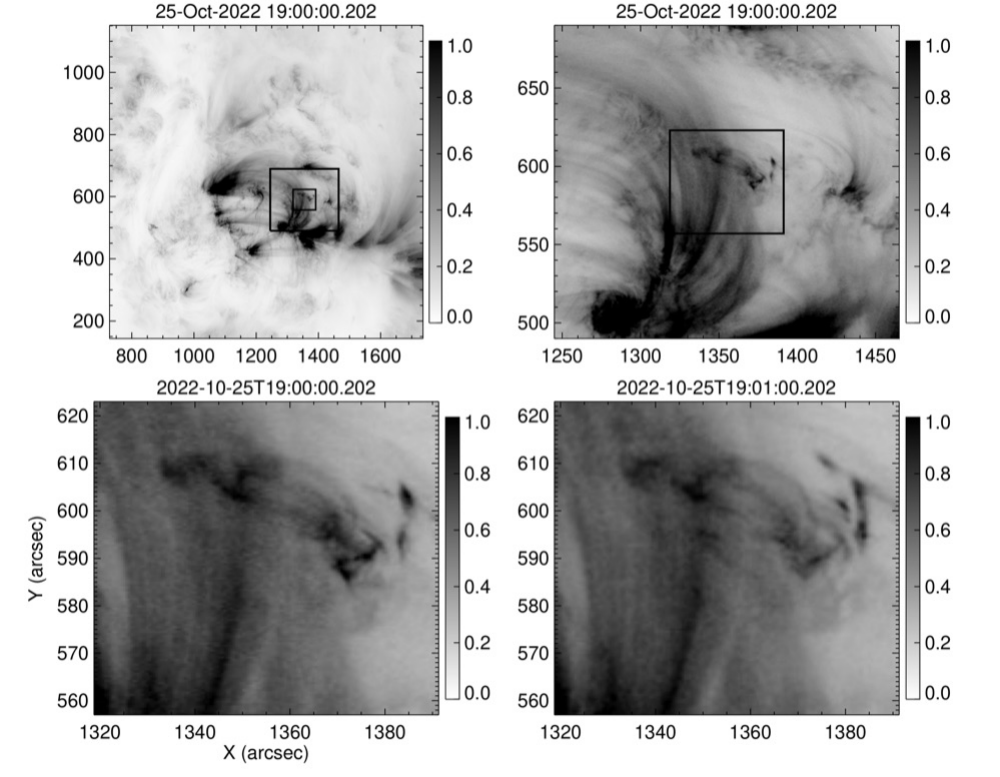} 
\caption{
\label{fig:one} 
The region of the Sun observed
by the EUI instrument is shown
as a full field-of-view (top left) and three close-up views. The lower two panels show the two types of structure
discussed in this paper, they are separated by just 60 seconds.  The large loops are
almost unchanged in 60 seconds.
However, there is a region of
dynamic evolution of EUV 
brightness in the upper half
of these two lower panels, possibly resulting from changes in magnetic topology. 
Intensities are on an arbitrary but 
identical scale for all panels. The squares show
the areas shown in the right and bottom  panels. 
}
\end{figure*}
}

\newcommand{\figtwo}
{\begin{figure*}
\includegraphics[width=\linewidth]{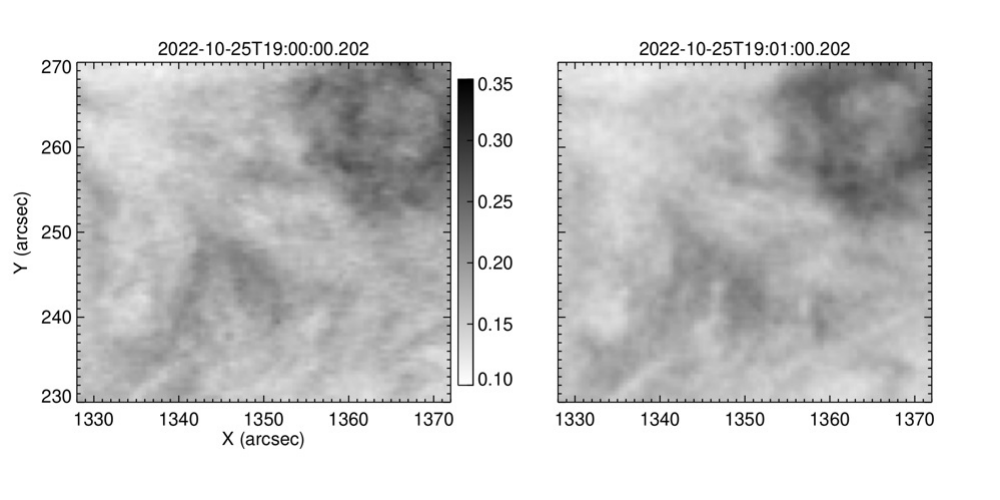} 
\vspace{-15mm}
\caption{
\label{fig:two} Close-ups of
a region of quiet Sun 
are shown for the same 
frames shown in Figure~\ref{fig:one}. }
\end{figure*}
}

\newcommand{\figthree}
{\begin{figure*}
\includegraphics[width=\linewidth]{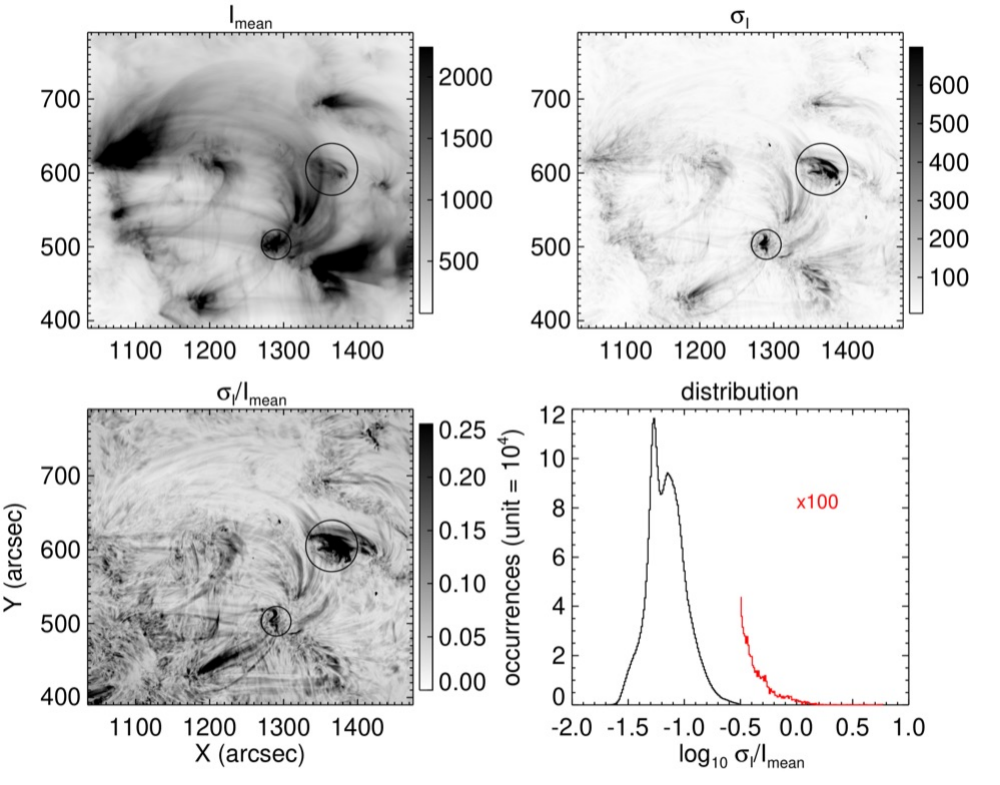} 
\caption{ The mean of the 
timeseries is shown (top left) with the rms variations 
(top right), and their ratio
(bottom left).   
The encircled region of
enhanced variances is also indicated in Figure~\ref{fig:sdo}. 
The line plot
shows the distribution of
$\log_{10} \sigma_I/I_{mean}$,
with the occurrences of larger variances 
magnified by a factor of 100
(red line).   Only 0.14\%{}
of all the pixels 
have values of 
$\log_{10} \sigma_I/I_{mean}$
exceeding 
0.1 in the logarithm, i.e. 
with values of $\sigma_I$ 
exceeding the mean by $26\%$.
\label{fig:three} }
\end{figure*}
}

\newcommand{\figpspec}
{\begin{figure*}
\includegraphics[width=\linewidth]{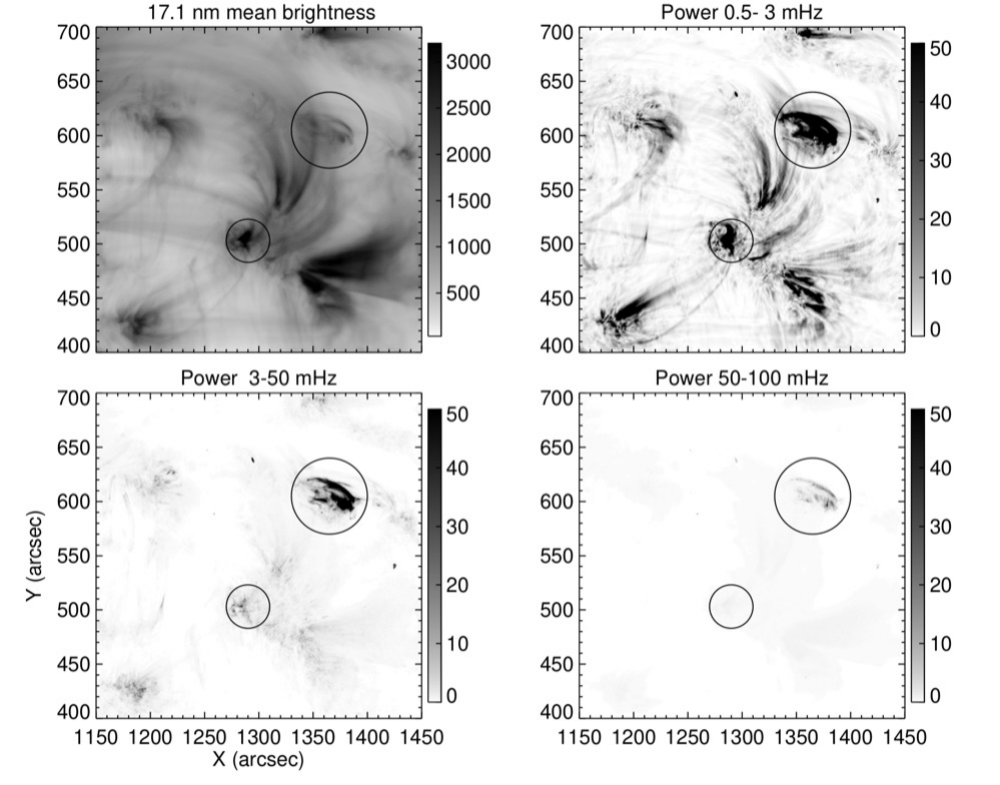} 
\caption{ Frequency-delimited  variances (integrated power spectra) of the 
EUI timeseries are shown as in Figure~\ref{fig:three}.   
\label{fig:pspec} }
\end{figure*}
}

\newcommand{\figsdo}
{\begin{figure*}
\includegraphics[width=\linewidth]{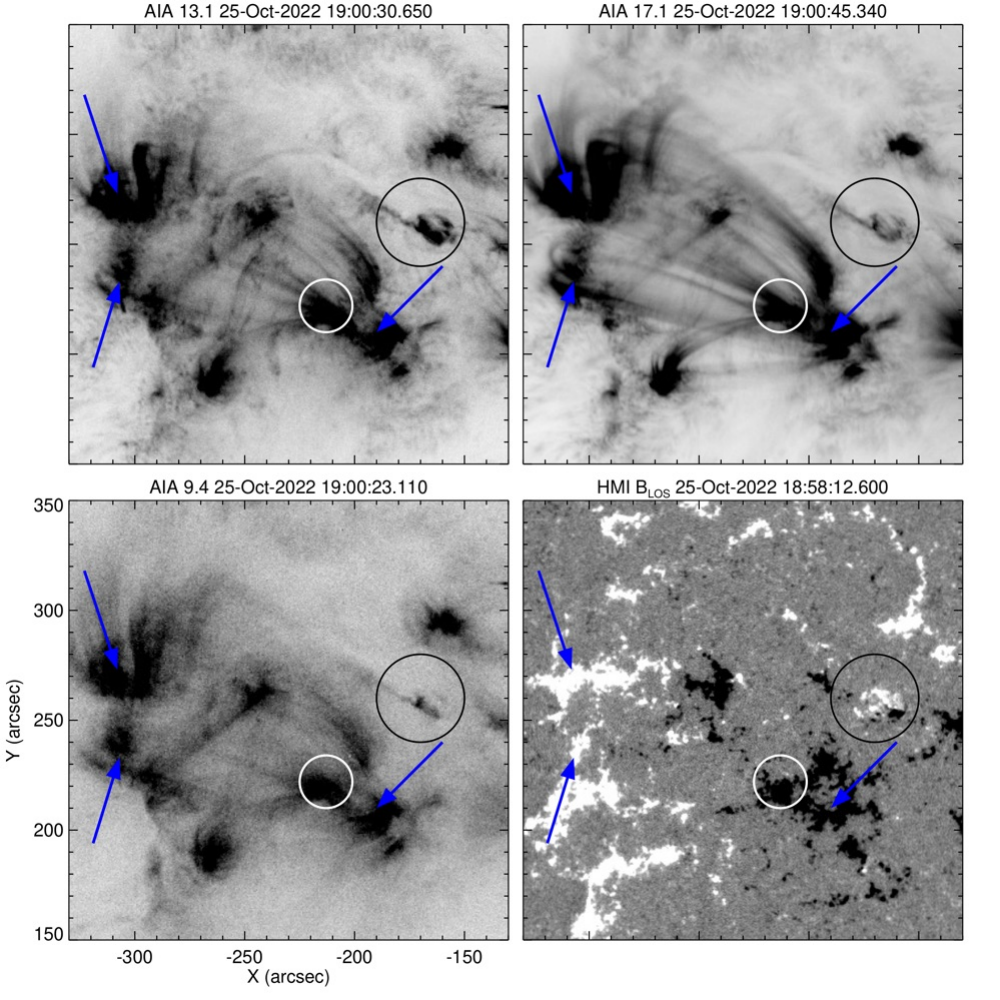} 
\caption{Four images 
observed from SDO, including the 
13.1, 17.1 nm and 9.4 nm AIA images which sample
plasmas near 0.5, 1 and 6 MK respectively, outside of flares. 
Arrows point to 
locations where the loop
system discussed in the text are anchored. 
 The close-up
views in Figure~\ref{fig:one}
are centered near $X=-180,Y=250$ 
in this figure. 
The encircled region
in black shows a magnetic bipole
emerging during 
the observations reported here.  
\label{fig:sdo} }
\end{figure*}
}

\shorttitle{}
\shortauthors{Judge}

\newcommand{\hao}{
High Altitude Observatory,
National Center for Atmospheric Research,
Boulder CO 80307-3000,
 USA}

\begin{document}

\title{Steadiness of coronal heating }


\correspondingauthor{P.  Judge}\affiliation{\hao}

\author{P.  Judge}\affiliation{\hao}

\date{Accepted . Received ; in original form }


%
%

\begin{abstract}
The EUI instrument on the Solar Orbiter spacecraft has obtained the most stable, high-resolution images of the solar corona
from its orbit with a perihelion near 0.4 AU. A sequence of 360 
images obtained at 17.1 nm, between
25-Oct-2022 19:00 and  19:30 UT is scrutinized. One image pixel corresponds to 
148 km at the solar surface.  The 
widely-held belief 
that the outer atmosphere of
the Sun is in a continuous 
state of magnetic turmoil is pitted against
the EUI data. 
The observed 
plasma variations appear  to fall into two classes.
By far the dominant 
behavior is a very low amplitude 
variation in brightness ($1\%{}$) in the coronal loops, with
larger variations in some footpoint regions. 
 No hints of observable 
changes in magnetic topology are associated with such small
variations. The larger amplitude, more rapid, rarer and 
less-well organized changes are associated with flux emergence. 
It is suggested therefore that 
while magnetic reconnection drives 
the latter, most of the 
active corona is heated with no evidence
of a role for large-scale (observable) reconnection.  
Since most coronal
emission line widths are subsonic, the bulk of coronal heating, if
driven by reconnection, can only be
of tangentially 
discontinuous magnetic fields, with
angles below about $0.5c_S/c_A \sim 
0.3\beta$, with $\beta$ the plasma beta parameter ($\sim 0.01)$, and $c_S$ and $c_A$ 
sound and Alfv\'en speeds.  If heated by multiple
small flare-like events, then these must be 
$\lesssim 10^{21}$ erg, i.e. pico-flares.  But 
processes other than reconnection have yet to
be ruled out, such as viscous 
dissipation, which 
may contribute to the steady heating of coronal loops over active regions. 
\end{abstract}


\section{Introduction}
\label{sec:statement}

The solar corona is commonly perceived to
be in a state of 
continual dynamic
evolution, as it responds to evolving magnetic
fields emerging from
the physical surface and
the more tenuous chromosphere.
Several years ago, of four central characteristics deemed
important in a review of coronal heating, we read
\citet{DeMoortel+Parnell2015}
\begin{quote}
    ``coronal heating is intrinsically non-steady.''
\end{quote}
This statement is not debated here.  Indeed, 
theoretical considerations imply that almost all mechanisms must be 
non-steady on scales of dissipation, which probably lie between ion viscous damping and kinetic scales (100 km to 10 m, e.g. \citealp{Davila1994}).  These mechanisms occur  below observational 
scales currently achieved. 
But, 
to what degree does our observational evidence require that the corona be intrinsically variable?  Observationally, dynamics is manifested in direct 
motions 
resolved in images, and through line profiles revealing unresolved motions. 
The latter have consistently revealed that the dominant 
unresolved dynamics is weak, subsonic
\citep{Billings1965, Bray+Loughhead+Durrant1984, Thomas+Neupert1994,
Hara+Ichimoto1999,
Raju+others2001,
Singh+others2002,
Kosugi+others2007,
Coyner+Davila2011,
Krishna+others2013,Brooks+Warren2016,Koutchmy+others2019}
with very occasional rapid 
motions associated with observable changes in apparent 
topology, i.e. magnetic 
reconnection.  But what about observations of
resolved motions?

The modern view  that the coronal is heated 
impulsively and dynamically has become 
so widespread as to be 
rarely questioned.
Of many papers 
the recent work
of 
\citet{Tiwari+others2023} is an example in which 
non-steady coronal  behavior is  actively sought.   One clear exception is found in a review, ``The dynamic solar corona in X-rays with Yohkoh'' by \citet{Tsuneta1996}:
\begin{quote}
    ``\ldots Yohkoh observations also show the existence of steadily heated plasmas with temperature of 2 - 4 MK, both in active regions and in the quiet Sun. The mechanism of the steady heating has not yet been understood.''
\end{quote}
In a related and prescient aticle, \citep{Sturrock1999}
questioned whether the 
observable magnetic fields are active or passive in heating the corona.  He reasoned that coronal heating may well result from 
processes well below the scales of our observations, i.e. there may be ``hidden variables'' in the coronal heating problem.   This idea appears to be untested, as yet. Indeed, 
influential  studies have sought direct relationships between 
some measurable magnetic field 
and coronal heating, with mixed success
\citep{Fisher+others1998,Mandrini+Demoulin+Klimchuk2000,Aschwanden2001}.
It is unclear if the 
inconclusive results 
result from Sturrock's
idea or if we simply 
have used data inadequate for
purpose.  

In viewing recent high-quality  movies 
of \citet{Auchere+others2023}
from
the Solar Orbiter EUI instrument \citep{Orbiter,Orbiter2, EUI, EUIc}, the present author
was struck that the 
dominant 
signal from the corona appeared to be 
constant.   A brief 
inspection of the EUI
data suggested also that observed  variations appear  to be attributable to 
benign, generally sub-sonic field-aligned flows, which are not 
necessarily associated with irreversible energy dissipation.  At the
small scales sampled
by EUI (148 km) during these observations, the only clear, rapid  ``activity'' usually 
associated with 
magnetic reconnection 
seemed to be associated with visibly 
emerging magnetic flux.

The consequences 
of a dominant mode of  apparent steadiness in 
the brightness of
the bulk of
coronal loops 
are intriguing, perhaps even vital
to identifying
the dominant heating mechanisms.   Reconnection, often associated with coronal heating
\citep[e.g.][]{Pontin+Priest2022}, appears to be absent most of the time in the data examined here, \textit{on observable scales.  }

This paper 
re-examines the concept that 
the  corona is observed to be in a state of
continual dynamic re-adjustment.  This 
idea has historically been inferred from decades of
space observations at X-ray, EUV and 
UV wavelengths, using 
evidence that is, when scrutinized,  largely indirect \citep[e.g.][]{Viall+others2021}.   Is the observable variability 
of coronal plasma  related directly to irreversible heating, such
as commonly assumed by,
for example, nanoflares, reconnection 
or wave turbulence?
Or is heating associated with physical processes which are unresolved 
in space and time? 
 To investigate this question, 
the primary data used below are a time series of 360 stabilized images at 17.1 nm from 
the Solar Orbiter EUI instrument obtained over
half an hour on 25th October 2022.

\section{A brief review of coronal dynamics and heating}

Many articles have derived 
quantities related to
coronal dynamics,  under the general
picture that the corona observed at X-ray and EUV 
wavelengths consists of
plasma loops.  
Section 2 of the review 
by \citet{Viall+others2021}
summarizes the literature 
on observational constraints on coronal heating, representing perhaps the 
generally accepted 
state of the subject.

Following \citet{Rosner+Tucker+Vaiana1978}, the current  paradigm within which 
most studies lie,  
is that the radiating coronal plasmas  represent bundles
of ``one-dimensional atmospheres'' owing to the 
high conductivity of coronal
plasma, the enormous physical
scales involved, and the frozen-field condition 
(Alfv\'en's theorem,
\citealp{Alfven1942}). 
One example
of this kind of motion in coronal plasma is described
by \citet{Peres2000}.
In an example of
confronting data with theory, 
Section 4 of  \citet{Schmelz+Winebarger2015} analyzed  lifetimes of active region plasma  loops in terms  of necessarily simplified models within
the accepted paradigm. 
Lifetimes 
were found to be significantly longer than those expected based on the calculated cooling times. This example reveals  recurring  general problems  in 
such data-model comparisons.  In order to make any progress 
the model simplifications necessarily include 
a host of assumptions which are 
justified neither by consideration of 
first principles, nor by
 interpretations
of available data which
are highly non-unique.

In a different approach, based upon 
some early hard X-ray observations 
\citep{Lin+others1984}
and 
a conflict between 
the plasma equations of motion and a fixed topology, 
\citet{Parker1988} 
proposed that bursts of 
nanoflares might heat the active corona.  In Parker's approach, 
basic theoretical results in highly
conducting plasmas 
were combined with 
available observations to propose that flares smaller than those reported by Lin and colleagues may naturally
supply a mechanism 
to deliver ordered 
magnetic energy into 
heat via the formation
of elementary current sheets \citep{Low2022}. 
Parker's approach has survived intense scrutiny and prompted a
significant community to
seek signatures of nanoflares \citep[e.g.][]{Bogachev+others2020}. 
Nanoflares have therefore been 
accepted as a likely 
candidate to explain heating of long-lived coronal structures
\citep[e.g.][]{Pontin+Priest2022}.  
However, the observational methods used to test the picture have, as above, contain elements which are  arguably  \textit{ad-hoc}, and  details of
any ``models'' are so
incomplete to make 
solid, testable predictions far from
our capabilities.  In short, this dynamic model remains 
almost untested by available
methods,

A recent monograph 
by \citet{Judge+Ionson2023}
digs deeper into 
assessing this current 
understanding.  
They highlighted 
many 
arguments 
why  the current paradigm from basic
theory should be 
re-examined, from basic theory 
to the limitations of remotely sensed data. 
It
is in this spirit that the current paper 
asks, using the unique
capabilities offered
by new capabilities,
if, and by how much, the corona is heated
steadily.  Readers
will perhaps be surprised that this
question is being asked in
2023, given the general 
tone within the current  literature.
For those we ask 
for patience as the novel
data are examined below.

\figone

\section{Observations}

We 
have analyzed  the 
sequence of images 
documented in Table~\ref{tab:obs}.   
\begin{table}{
\caption{EUI observations of the corona on 25th October 2022}
\begin{tabular}{lll}
\hline\hline
Times & 19:00:00.202 & 19:29:55.205\\
Format & $2048\times2048$ \\
Distance to Sun & 60,755,600 km & 0.406125 A.U. \\
Number of frames  & 360\\
Cadence & 5 seconds\\
Pixel size & $0.492\arcsec$ & 148 km\\
Bandpass & 17.1 nm \\
\hline
\end{tabular} 
\label{tab:obs} }
\end{table}
The only processing 
performed beyond
the publicly released EUI data was
to perform an accurate co-alignment
of the time series. 
All images were aligned to the time-averaged image using cross-correlations. 
A new average was computed and the process
repeated once.  The residual jitter is less than 1 pixel
($\equiv 148$ km at the solar surface). 
These data are comparable to the highest resolution EUV images
obtained, with the HiC suborbital
rocket experiment \citep{Rachmeler+2022}.  HiC obtained images at 17.1 nm 
with a plate scale of
0.129\arcsec, which is about 94 
km, compared with 148 km in Table~\ref{tab:obs}.   The spatial  resolution of 
HiC data assessed by \citet{Rachmeler+2022} is complicated
by time-dependent jitter, instrument PSF and other properties of the images obtained.  The authors concluded
that only a fraction of their 5 minutes of data was unaffected by
jitter, yielding a resolution
$> 2\times$ the pixel size or worse.  The 
shorter duration of the sequence of HiC images 
are less suitable for our explicit goal of studying dynamic evolution of the coronal plasma.

Most of the data analyzed by 
\citet{Rachmeler+2022}
achieved widths (FWHM) across
coronal strands 0.4\arcsec or greater
(290 km), essentially  the same physical
size as the Nyqvist-limited  sampling 
resolution of the EUI images.   While these different datasets have similar
imaging performance, the Orbiter
data span a  longer time period. 
 Significantly, the 30 minutes of
data is longer than the 
typical cooling time
of coronal loops
in active regions
of a few minutes.

Figure~\ref{fig:one} shows
the field-of-view and 
close-ups.   The two bottom
panels 
highlight the same region 
separated by 60 seconds, to emphasize the two different types
of behavior of concern to the present work.  ``Constant'' long coronal loops
don't change perceptibly between these two panels.   But 
a region of more broken-up
emission evolves more dramatically 
on short timescales.   It would be difficult to identify field-aligned motions 
in the latter structure, seen ``underneath'' the more obvious static loops.

A third type of
morphological  structure 
is shown in 
Figure~\ref{fig:two}, an area
of quieter Sun from these same images which is closer to the solar equator.
This area shows few, if any,  clear loop structures. The diffuse emission 
does vary on timescales of
60 seconds, but the morphology of the structures is 
not as clear as those
highlighted in Figure~\ref{fig:one}.  The active region 
data analyzed here are sufficient to 
address an important aspect of the main  question of 
the nature of
coronal heating.  The quiet regions 
will yield different outcomes, but the magnetic lines  of force cannot arguable be traced, presenting
a challenge to 
assigning variations as field-aligned or 
otherwise.  

\figtwo

\subsection{Statistics}

Figure~\ref{fig:three}
shows the simplest possible
statistic of variation, 
i.e. those independently calculated 
for each of the $2048^2$ pixels. 
The figure shows just the central (most active) part of the EUI field-of-view, to highlight
coronal loops. This
area lies between
those shown in the upper panels of
Figure~\ref{fig:one}.

For each
pixel we  evaluated 
the variance $\sigma_I^2$ 
from the time series.  The image
of $\sigma_I$ 
results almost entirely  from solar, not instrumental  changes. As noted 
above, this image is very different 
from an image of $\sqrt{I_{mean}}$,
which is similar to the image of $I_{mean}$ itself. 

Aside from
several, rare ``hot spots'' of activity
(examples are circled in Fig.~\ref{fig:three}),
the variations as measured
merely by $\sigma_I/I_{mean}$ 
lie mostly below 0.1 in
the logarithm.   The 
distribution of these variations is shown in
the bottom right figure. 
Above this value, which  corresponds to 
$\sigma_I/I_{mean} \ge  10^{0.1} = 0.26$, less than 
1\% of the pixels show
larger variations,   The vast bulk of 
the variations lie near
 $10^{-1.3}I_{mean}$.  The most likely variation in the time series thus corresponds to
$$
\sigma_I = 0.05 I_{mean}.$$  The large loop
system in the figure has  $\sigma_I/I_{mean} \sim 0.01-0.02$,
which may be a lower limit if
coronal motions 
such as sub-pixel oscillations 
on the plane of the sky reduce 
variances.  
The  more highly variable
regions marked in red are spatially associated
with the circled 
nests of activity.
Figure~\ref{fig:pspec} shows  variances split over various frequency ranges.
\figthree
Although these figures show 
just local behavior, it is clear
that there is a great
deal of pixel-to-pixel structure in these quantities.  Notice in particular
the large loop systems in $I_{mean}$ and their unusually small variances along the bulk of the loop
lengths.   When loop-like 
variances are larger,  they trace out loci which are along 
plasma loops observed only in the intensity images.
Occam's Razor would
suggest the  
interpretation that 
field-aligned
mass, momentum and energy transport, and/or transverse loop oscillations 
must be responsible for this
observation.     There is no obvious need to
suggest other sources of
variation, but this might be explored
in future (section~\ref{sec:future}). 

In stark contrast, 
the smaller-scale
nests
of large variances
are clearly
seen as 
amorphous shapes. Two examples  
are circled in 
Figure~\ref{fig:three}.   
The nests appear 
to be examples of
a ``geyser'' phenomenon 
(\citealp{Paraschiv+Donea2019}, A. Paraschiv private communication 2023).

Figure~\ref{fig:pspec} shows
power in ranges from 
0.5-3 mHz, 3-50 mHz, and 50-100
mHz.  The only structures 
with power over all frequencies
are associated with the nest near
$X=1365,Y=605$, which we examine next.

\figpspec

\subsection{Signatures of emerging magnetic flux}

Contemporaneous data from both the HMI and 
AIA instruments on SDO were examined. Examples are
shown in Figure~\ref{fig:sdo}, over which 
the two encircled regions 
of Solar Orbiter coronal data of  Figure~\ref{fig:three} 
are also indicated.  Both 
show 
emergence of flux of opposite polarity. The larger circled
region  at $X=-160,Y=255$ in SDO coordinates  
emerged  
$\approx 5\cdot 10^{19}$ Mx of opposite polarity flux over 2 hours from
15:00 UT.   This
is typical
for this size of emerging region,
as observed with HMI \citep{Norton2017}.  
 The other (near $X=-220,Y=220$)
has significantly less emergence of flux of opposite-sign to the dominant polarity. Its behavior typifies other regions nearby.

We might use the images of variances in 
Figure~\ref{fig:three}
to identify coronal structures.  The first and 
most obvious is that the 
loops are readily apparent in images of $\sigma_I$, especially near the footpoints 
(marked with arrows later in Figure~\ref{fig:sdo}).   
The amplitudes 
of the loops visible in 
$\sigma_I/I_{mean}$  in 
Figure~\ref{fig:three}
are almost all below 
0.25.  Strikingly, the longest
loops in the top left of each panel show particularly
small variances
below 0.05 in $\sigma_I/I_{mean}$. The bright coronal structure 
near $X=1400,Y=500$ has
variances which are similarly suppressed. 
Lastly, those structures 
with the largest variances 
are geometrically far smaller than the 
typical active region loops.

\figsdo

The SDO data shown in Figure~\ref{fig:sdo} highlight 
the central part of the 
field of view  observed by 
EUI on Solar Orbiter.   The
plasma loops observed at 1 MK (at 17.1 nm
wavelengths) sparsely connect regions marked with arrows 
The hotter plasma at 9.4 nm 
(mostly from \ion{Fe}{18} formed near 6 MK) is more diffuse.  With the co-spatial 
13.1 nm emission formed
near 0.5 MK, it is clear that 
this loop system contains a broad distribution of plasma
with respect to electron temperature.
The 
 bipolar emerging flux region identified
by the black circle, exhibits
smaller, less smooth  coronal structures
with increasing temperatures, the opposite of the observed  loops.

\section{Discussion}

This work addresses a very simple question: what does
the apparent stability of coronal structure tell us 
about underlying plasma 
heating mechanisms? 
With the advent of
excellent data from the EUI instrument of
Solar Orbiter, the analysis is accordingly very simple, and as such, stands in
contrast to some of the vast literature on
related work, which is considerably more complicated, and laden 
with many more assumptions
\citet[see the recent review of ][]{Viall+others2021}.
Nevertheless, the
present work suggests that 
significant information has been 
overlooked for decades.  Sometimes
in nature it is those data which do not immediately attract our attention
which are most important.   In this particular case emphasis is given
to the \textit{lack}
of activity in particular within 
typical 
coronal loops in
active regions. It was
concluded above that the vast bulk of  solar coronal plasma over active regions 
is heated quiescently, when observed on scales down
to 150 km.    The radically different   behavior
at 17.1 nm emission 
between coronal loops
and regions of  emerging flux has
been quantified in
various figures.  Most notably, the power spectra are heavily
weighted towards 
low frequencies
(0-3 mHz) in loops,
higher frequencies
above emergent regions (Figures~\ref{fig:three} and \ref{fig:pspec}). Further, the amplitudes
of variation are clearly reduced along larger,
long-lived loop systems (see the upper half of
the lower left panel of Figure~\ref{fig:three}).  Also,
variances are larger in some regions close to loop footpoints, but not all.   

\subsection{Possible implications}

Whatever the dominant mode of coronal heating of plasmas near and above 1 MK, it must 
respect the new basic observational
constraints presented here, and  the well-documented subsonic coronal
emission line widths
and shifts.
 In such conditions, the only significant  irreversible fluid effects are  energy losses by radiation and an increase of entropy by field-aligned heat conduction, but with no
genuine source of irreversible heating, such as shocks.     Those
near-footpoint regions showing a larger variance than the looptops
would appear
to correspond to
such motions, given
the literature 
on the dynamics of
loop footpoints
\citep[again one can find a discussion in ][]{Viall+others2021}.
In spite of the emphasis on supersonic motions
with speeds $> 100 $ km~s$^{-1}$ in the literature  
\citep[e.g.][]{dePontieu+others2009roots,Raouafi+others2023}, such motions appear largely absent in some typical active region loops, the vast majority of the time.  

If   
irreversible heating 
cannot occur through shocks, it must be through
another kind of
dissipative structure, associated with 
electrodynamics, 
hydrodynamic processes having no further 
avenue for dissipation. Reconnection produces flow speeds of order the Alfv\'en speed of the oppositely
oriented components of the magnetic field.   With magnetic field strengths of order 
$B \sim 100$ G and 
number densities of
protons close to 
$5\cdot 10^9$ 
cm$^{-3}$ for
active region loops,
the Alfv\'en speed
$c_A$ is $ \sim 2800$
km~s$^{-1}$. The linewidths therefore
suggest that 
reconnection, if important, must involve tangential
components of the magnetic field with a strength $ \leq 1$ G, accompanied by flows of $\leq 30$ km~s$^{-1}$.   In other words, only tiny fractions of the total magnetic field in the corona must be responsible for coronal heating. 
Interestingly, \citet{Sturrock1999}
already 
wondered if the 
measurable magnetic fields defining the coronal topology play an active or passive role in 
coronal heating.

It is widely recognized that viable heating 
mechanisms intrinsically involve departures
from the symmetry implied 
in a 1-dimensional picture, even if
the fluid is required to flow
within such structures \citep[e.g.][]{DeMoortel+Parnell2015}.
In the
presence of asymmetries
within what appear to be 
one-dimensional plasma loops, 
dissipative structures develop through spontaneous 
current sheet formation, and 
intermittent 
phase mixing, resonant absorption, and 
of internal surface waves. These processes dissipate the ordered energy of velocity gradients, manifested in bulk flows of ions and electrons and 
differential flows (electric currents), 
through  
particle
collisions, the final step in  
irreversible plasma heating.   These particular processes are familiarly called 
viscous and Joule energy  dissipation
respectively.  
These sources of heat are
absent in flows which are essentially one-dimensional, within
strong fields which
are close to potential, with weakly compressive motions \citep[e.g.][]{Hollweg1986}. 
In this regard, 
on resolvable scales, active regions are frequently surprisingly close to potential fields 
\citep{Wiegelmann+Sakurai2021}.  
This appears to be 
true for the longer
loops 
showing small variances analyzed here \citep{Schrijver+others2005}.

\subsection{On the role of magnetic reconnection}

Let us examine the hypothesis that 
coronal loops 
are heated by processes originating from magnetic reconnection. 
The word \textit{originating}
is important, for there 
is a vast literature on
the role of reconnection
following a cascade of
turbulent motions to 
unobservable scales 
\citep[Appendix D of][]{Schekochihin2022}.  This second kind of
reconnection may well be present, but it is the result of a different 
primary process, occurring on small scales, and 
its observational signatures may be hidden
below our ability to measure.   

To proceed, 
in the kind of
loops observed by EUI, 
some typical conditions can
be estimated from many previous studies \citep[e.g.][]{Jordan1992}.
The Alfv\'en speed
is estimated to be  
\begin{equation}
c_A = \frac{B}{\sqrt{4\pi\rho}} \approx 2800\ 
\frac{B_{100}}{\sqrt{ n_{9.7}}} \ \ \mathrm{km~s^{-1}} 
\end{equation} where $B_{100} \sim 1$ 
is the magnetic field strength in units of
100 G, and $n_{9.7} \sim 1$ is the number density of coronal ions  
in units of $5\cdot 10^9$ particles
per cm$^{3}$.   

Magnetic reconnection
is first and foremost
a mechanism for a 
magnetoplasma system to relax rapidly to
a lower energy state via 
a change in magnetic topology. In doing so, newly unbalanced Lorentz forces  
accelerate plasma 
to drive bulk plasma motions to form
the exhaust of a small-scale magnetic diffusion region.  Other \textit{observed} effects of reconnection as 
witnessed during flares and 
plasmoid ejections 
include 
particle acceleration, hard X-ray 
emission and plasma turbulence
\citep[e.g.][]{Fletcher+others2011}. It can be a source of high-frequency waves with
additional consequences
\citep{Kasper+others2013}. But these 
are secondary effects 
in the highly conducting, large 
low$-\beta$ corona.  They may or may not play an active role in
heating the corona.

The speed with which the 
bulk plasma is swept up
by reconnection approaches  the Alfv\'en speed associated with the reconnected vector component of
the total magnetic field, $
\mathbf{B_\perp}$, where
the component along the 
loop diection is $\mathbf{B_\parallel}$,
$$
\mathbf{B} = \mathbf{B_\parallel} +
\mathbf{B_\perp}.
$$
The weakly varying intensity
data from EUI,  and subsonic linewidths, would seem to confine the possible role of magnetic reconnection to $\mathbf{B_\perp}$, with 
$\mathbf{B_\parallel}$ playing just a  passive role, 
as follows.   With 
$\xi\lesssim 30$~km~s$^{-1}$ as an 
estimate of the Alfv\'en speed 
of the components
of reconnecting magnetic fields, $\mathbf{B_\perp}$, then we can 
estimate that 
\begin{equation}
B_{\perp} \lesssim \ 
  \xi_{30} \ \sqrt{n_{9.7}},
\end{equation}
where $\xi_{30}$ measures the linewidths  
in units of 30 km~s$^{-1}$.
Thus, 
if these motions are 
caused by annihilation of 
$\mathbf{B_\perp}$, 
then its magnitude is
$\approx 1$ G,  1\%{} of the  magnetic field
strength $B$ expected under typical
conditions 
\citep[e.g.][]{Jordan1992}, also  
noting the photospheric field
strengths of
Figure~\ref{fig:sdo}.
In the 
``nanoflare'' picture of \citet{Parker1988},  ratio $B_\perp/B$ 
 equals the angle 
between  magnetic fields separated
by tangential discontinuities.  
\citet{Parker1988} estimated  
a much larger angle of 0.25 radians.  This factor of 25 
discrepancy will be 
explored below. 

Further  implications from
the EUI data on reconnection are  
debatable, because 
imaging, even at ``small'' scales of 150 km, is unable to measure
scales at which dissipation 
must occur, except for a possible role for
ion viscous dissipation
near $10^2$ km
\citep{Hollweg1986,Davila1987}.  Undaunted by this, we
follow arguments using 
decades of observations 
of EUV observations
\citep[reviewed by][]{Bogachev+others2020}, 
and explore quantitatively what the 
findings from EUI imply
in terms of elementary
units 
(i.e. ``quanta'') of energy
dissipation. 

Let us assume that
all the brightness variations 
observed are due to 
reconnection somewhere within
a plasma loop, then Figures~\ref{fig:three} and 
\ref{fig:pspec}
indicate that the majority of
pixels within loops have 
$\sigma_I/I_{mean}$ of order 0.01 (see the 
lower left panel of Figure~\ref{fig:three}), with occasional fractions rising 
to 0.25 near specific footpoints.   Assume further
that each such pixel 
is connected by rapid heat conduction along the length
$L$ of the bundle of 
flux, so that the 1\%{} 
variations apply to the entire volume
\begin{equation}
{v} \approx L p^2
    \end{equation}
where $p=150$ km is 
the EUI pixel size. Within 
volume $v$ let
$\mathcal{E}$ erg be the energy released in each 
reconnection event.  The 
total rate of energy release
within volume $v$ is 
\begin{equation}
    \dot n \, \mathcal{E}
    \mathrm{\ \  erg~s^{-1}}
\end{equation}
or a volumetric heating rate of 
\begin{equation}
    \frac{\dot n\,\mathcal{E}}{v}
    \mathrm{\ \  erg~cm^{-3} s^{-1}}.
\end{equation}
Here, $\dot n$ is the number of events 
of energy $\mathcal{E}$ per unit time.  Our 
immediate aim is
to find values of $\mathcal{E}$ and $\dot n$  from the statistical variations in the EUI data of loops, using known energy flux
density requirements. 
From previous
observations \citep{Withbroe+Noyes1977}, we know that for a loop of length $L$, the energy flux densities 
$W_{obs}=10^7$ erg~cm$^{-2}$s$^{-1}$, so that 
\begin{eqnarray}
    \frac{\dot n\,\mathcal{E}}{v} &\approx& \frac{W_{obs}}{L} \ \ \ \mathrm{erg~cm^{-3}s^{-1},~and} \\
    \mathcal{E} &=& \frac{W_{obs} p^2}{\dot n} \ \ \mathrm{erg.}
\end{eqnarray}
To estimate $\dot n$, we use 
the EUI observation that
$\sigma_I\sim 0.01 I_{mean}$,
for the bulk of the loop system seen in
the figures. 
When $I_{mean}$ and $\sigma_I$ result from statistical fluctuations of $n$ quanta of energy, then $1/\sqrt{n} \sim \sigma_I/I_{mean}$.  Then $n \sim (I_{mean}/\sigma_I)^2$.  For variations of 0.01,
we find 
$\dot n\approx 10^4/1800$
quanta per second, 
from which 
\begin{equation} \label{eq:eest}
    \mathcal{E} \approx 4\cdot 10^{20} \ 
    \mathrm {erg},
\end{equation} three to four orders of
magnitude smaller than Parker's original 1988
estimate.   Note that $\mathcal{E}$ varies as the area of each 
instrument's pixel.  If larger variations of order 0.1
are present in, say, the 28.4 nm band, then 
$\dot n\sim 100/1800$ quanta per second, 100 times smaller,
and the energy in each event would be $\mathcal{E}\approx 4\cdot 10^{22}$ erg. 

Alternatively if $\mathcal{E}$ were
instead $10^{24}$ erg (a ``nano-flare'', 
Parker 1988), we would
expect to see changes 
of order unity along the loops in the EUI images.
Or, for a pixel size of say $p= 725$ km ($1\arcsec$ from
Earth's orbit), the rms  variations would be reduced by the factor
$725/150 \approx 5$. For the $725$ km pixels, fractional brightness  variations would be 
reduced to 0.2.

The far smaller 
estimate of the energy of quanta given by equation~(\ref{eq:eest})
results from both
the small pixel areas $p^2$
available from the EUI on
Solar Orbiter, as well as
the tiny 1-2\%{} rms 
changes of intensity measured along these loops. 
Of course, these results explicitly assume energy released in quanta, and 
they
must be upper limits because
other sources of brightness variations, such as 
waves, flows, 
have been ignored.  Next we
try to reconcile the
divergence of the present 
results from those found
earlier. 

\subsection{Nanoflares revisited}

Parker's 1988 estimate
of the properties of
nano-flares with energies 
of order $10^{24}$ erg
was originally derived using two
sources of data:
\begin{enumerate}
    \item The average energy loss of the active coronal plasma is $W_{obs}=10^7$ 
    erg~cm$^{-2}$~s$^{-1}$ from \citet{Withbroe+Noyes1977}.
    \item This is supplied by work done by convectively-driven photospheric plasma on 
    magnetic fields emerging upwards into the active corona.  He adopted 
    thermodynamic parameters from observations of granules, lifetimes $\tau$ of 500 seconds and horizontal
    (random) velocities of $u=0.5$  km~s$^{-1}$.   
\end{enumerate}

\subsubsection{Parker's original 
formulation}
The work $W$ per unit time done by horizontal granular motions
moving vertical magnetic fibrils of field strength $B$ at a speed $u$ 
against the magnetic tension force is:
\begin{equation}
    W \approx \frac{B_\perp B}{4\pi} u \ \ \ \  \mathrm{erg~cm^{-2}~s^{-1}}.
\end{equation}
Assuming from the ideal induction equation that
\begin{equation} \label{eq:bperp}
    B_\perp \approx B \tan \theta = B \frac{ut}{L}, 
\end{equation}
where $\theta$  is 
 the average angle between magnetic field vectors on
either side of a tangential discontinuity,
then 
\begin{equation} 
W \approx
  \frac{(uB)^2}{4\pi L}\  t  \ \ \ \ \mathrm{erg~cm^{-2}~s^{-1}}.
\label{eq:W}
\end{equation} 
With $B=100$ G, and for
a loop of length $L=100$ Mm, 
the time needed 
to generate $W_{obs}$
is $t=T\sim 5\cdot 10^4$ seconds $\approx 100\tau$, $\theta \approx 0.25$ radians, 
the total footpoint displacement $uT \approx 25 $ Mm, 
and $B_\perp \sim 0.25 B$. The total
footpoint displacement must occur
through the addition of $m \approx 100$ increments of length $u\tau$. For each  increment the energy released by reconnection across
the current sheet
is
\begin{equation} \label{eq:eparker}
\mathcal{E} \approx \frac{B_\perp^2}{8\pi}  \mathcal{V}   \approx 10^{24} \ \ \ \mathrm{erg}
\end{equation}
where the volume 
$\mathcal{V} \approx (u\tau)^2\Delta L$  cm$^3$, with 
$\Delta L = L/m \approx 1000$ km  is the extension along the coronal loop
of each current sheet element. The volume $\mathcal{V}$ is product of the $\Delta L$,  $u\tau$ and  
the width of the flux bundle, 
also assumed to be $\approx u\tau =250$ km.

\subsubsection{A revision of Parker's scenario}

Another Parker-like 
estimate of $\mathcal{E}$  comes from integrating equation
(\ref{eq:W}) in time. Let us assume that energy is released
in quanta after time $t=T$ with energy $\mathcal{E}$, and solve for these quantities given
Parker's parameters. 
A random walk with $m > 1$ steps of
duration $\tau$, each of length $u\tau < L$, yields  
\begin{equation} \label{eq:bperpq}
    B_\perp \approx B \tan \theta = B \frac{\sqrt{m}\, u\tau}{L}, 
\end{equation}
so that in time $T= m\tau$
\begin{equation} 
W \approx
  \frac{(uB)^2}{4\pi L}
    \sqrt{m}\tau
  \ \ \ \ \ \mathrm{erg~cm^{-2}~s^{-1}}.
\label{eq:Wprime}
\end{equation} 
With $W=W_{obs}$,  
$\sqrt{m}\tau = T=5\cdot10^4$ seconds as above, we find  instead 
$m\approx 10^4$.  The path length over which work is done
is $\sqrt{m} u\tau $ instead of
Parker's $mu\tau=T$, so that 
each 
quantum releases
\begin{eqnarray}
\label{eq:eintegrate}
 &\approx& 
\frac{(uB)^2}{8\pi L}\, T^2
\ \ \ \mathrm{erg~cm^{-2}, \ and~so}\\
\mathcal{E} &\approx& 
\frac{(uB)^2}{8\pi L}\, \mathcal{A}\, T^2
\ \ \ \mathrm{erg}.
\end{eqnarray}
The net area swept out by the moving bundle is 
$\mathcal{A}\approx mu\tau\cdot \ell$ where $\ell$ is the 
characteristic size length of the flux bundle perpendicular to $\mathbf{u}$
and $\mathbf{B}$. Using $\ell\approx u\tau$, then
\begin{equation}
    \mathcal{E} \approx 10^{21}\ \ \ 
 \mathrm{erg},
\end{equation}
three orders of magnitude smaller than
Parker's original estimate. 

\subsubsection{Further constraints}
 
With knowledge of
mass densities $\rho$ associated with the observable 
unresolved motions of amplitude $\xi$, whether  waves or ensembles of reconnection
jets, the energy flux $W$
can be estimated through
\begin{equation}
W \lessapprox \rho \, \xi^2 \, c
\end{equation}
with $c$ the group speed of the 
waves and/or speed of ordered 
plasma motions in jets. 
The inequality must be used because the above estimate assumes all energy propagates upwards, 
and becomes dissipated within coronal plasma along the magnetic flux bundle.
Combining 
the above estimates of $\rho$, $\xi$ and using
$c=c_A$, we find 
\begin{equation} \label{eq:fluxest}
W \lessapprox  3\ 10^7 \sqrt{n_{9.7}} \ \xi_{30}^2 \ B_{100}. 
\end{equation}
Independent of 
estimates of energy fluxes
from nanoflares,  $W_{obs}$ 
requires line widths $\xi$ of order  
$30$ km~s$^{-1}$.

Observed coronal line widths 
are $\xi \sim 30$ km~s$^{-1}$ \citep[e.g.][]{delZanna+Mason2018}, 
corresponding to $\xi \sim  0.2 c_S$
with
$c_S$ the sound coronal speed. 
Using as a rough estimate to satisfy the inequality above, 
$\xi \sim0.5c_S$, these speeds can
be identified 
with the Alfv\'en speed 
of the annihilated component
$B_\perp$ in Parker's picture. Then $B_\perp \sim B 0.5 c_S/c_A$, or $\tan \theta 
\sim 0.5 c_S/c_A \sim 0.3 \beta$, where the plasma $\beta = 8\pi p/B^2 \sim 0.01$.

Thus we arrive at the independent
estimate for the annihilated magnetic component $B_\perp
\approx 0.01 B$ using 
equation~(\ref{eq:fluxest}). This should be 
contrasted with  $B_\perp \approx 0.25 B$ as suggested by Parker.

In summary, it is proposed
that Parker's picture, if it is
indeed the source of heating of the active corona, be
quantitatively modified such that the nano-flares are more pico-flares with energies $\lesssim 10^{21}$ erg.
The tangential discontinuities 
are accordingly weak (misaligment angles of
$\sim0.01$ radians). Any reconnection of
the perpendicular components occurs 
rapidly at lower alignment angles 
and electric current densities than derived by
Parker.   It is quite possible, in the absence of data to the contrary, that 
processes other than reconnection may 
cause the apparently steady heating rates inferred from the new EUI data,
such as dissipative surface 
waves \citep{Ionson1978}, 
dissipation of compressive 
ion motions \citep{Hollweg1986,Davila1987},
MHD turbulence \citep{Rappazzo+others2007,Rappazzo+others2008,Einaudi+others2021}
and wave dynamics in inhomogeneous conditions 
\citep{Howson+others2020}.
The 3D numerical MHD experiments 
of \citet{Einaudi+others2021},
including stratification and treatments of thermodynamics, independently arrived at energies
between $10^{18}$ and $10^{21}$ erg
for their ``elementary events''.  Although MHD may not be 
applicable across the range
of scales leading to 
irreversible dissipation, in 
particular at kinetic scales 
(mean free paths are a few hundred km), \citet{Einaudi+others2021} argued that the total energies of elementary 
events is independent of magnetic Reynolds number. The convergence of
these two results is interesting.

Lastly, the number of pico-flares
per unit area per unit time is
$$
\frac{n}{Lp t} \sim 4\cdot 10^{-17} \ \ \mathrm{cm^{-2}~s^{-1}}
$$
This lies within the errors in the relation
$\log_{10}N(E) = 30.6 - (2.18\pm 0.2)\log_{10}E$,
of the form $N(E) \propto E^{-\alpha}$, 
values derived from 
SDO data \citep{Ulyanov+others2019}.  When  $\alpha > 2$ the heating is dominated by small events.   Thus
pico-flares, if indeed are responsible for
the EUI variability, are a dominant 
contributor to coronal heating in active regions.

\subsection{Consequences}

On scales down to 150 km, the solar corona 
observed by EUI at 17.1 nm exhibits 
essentially two kinds 
of phenomena.   The first
 dominates 
almost all of the active corona
almost all the time.   While not a unique
interpretation, the idea of quanta of
small flares has been extended to 
pico-flare scales ($10^{21}$ erg) enabled through
the almost unprecedented spatial sampling of
$p=150$ km, the very small variances, and the long
time series (1800 seconds) of the EUI observations.
Other physical mechanisms than small flaring 
may prove to be entirely compatible with the 
EUI data, no attempt to refute 
them is made here.  In one sense, the ``nano-flare''
picture of Parker remains intact, but at far smaller
scales.  

The second kind appears to
be connected with 
more explosive behavior
which has drawn 
the attention of decades of
solar observers.   We have no
measurements of magnetic field
changes, just changes in
image morphology, to support our claim. So at this
stage it remains a hypothesis to be tested.
But the consequences
are pivotal in our
quest
to understand the 
solar corona.  A host of
models for bulk coronal heating based upon
reconnection must now
be challenged. For example,
models by \citet{Priest+others2002},
drawing on the interaction of convection-driven 
multipolar fields on scales
of Mm and above 
must be called into
question.  This particular 
model, as well
as all models based 
on multipolar magnetic fields \citep{Dowdy+Rabin+Moore1986,Antiochos+Noci1986,Hansteen+others2014}
also fails to capture elementary
properties of the hot plasmas in relation to 
chromospheric magnetic fields 
measured recently 
with DKIST
\citep{Judge+others2023}.

Lastly, we note that 
the well-studied 
phenomenon of coronal
rain, sensitive to
excess heating events, 
fills less than about 3\%{}
of the active corona by volume
\citep{Antolin+others2015}.
This might suggest, again,
that the bulk of the time
the corona is in a state
where excessive heating events are rare,
and the corona appears to be steadily heated
on very small scales for long periods of time
(1800 seconds), longer than loop cooling times.

\subsection{Conflicts with earlier work?}

These results appear to be
in direct conflict with
much earlier work, yet arguably
they are based upon the 
most stable set of coronal
data ever obtained, at the highest angular resolution.
These data are also 
compatible with 
the decades-long 
set of measurements of
unresolved motions in
the ``undisturbed'' corona. 
Here, an attempt to
explain this discrepancy 
is made through examining 
the methods used elsewhere
which have delivered
a different kind of
understanding.

Firstly, 
Parker's (1988) 
article on nanoflare heating was prompted 
by full-Sun 
observations of hard X-rays by \citet{Lin+others1984}.
The X-ray data 
on 27 June 1980
were from 
a day close to 
the sunspot maximum
with at least 15 intense bipolar regions 
on the disk, as judged
from the Kitt Peak 
Vacuum Telescope 
magnetograms acquired
around 18:00 UT (as stored 
in the Virtual Solar Observatory archive). 
Lin et al. 
found that hard X-ray bursts ($>$ 20 keV) occurred about one every 5 minutes above a flux of $\sim 7\ 10^3$   cm$^{-1}$ s$^{-1}$ keV$^{-1}$).  They speculated that the energy of the accelerated electrons might be comparable to that required to heat the active corona.   However, no 
sources were identified 
on the Sun in these
full-Sun measurements. 
The relation of this early work to the data
presented here are at best, questionable.   There is clearly no direct conflict 
between these two sets
of data. 

A second tentative conflict arises considering  observations of transition-region plasma (here we use
a definition to include all
plasma emitting from the Sun
between electron temperatures
of 20,000 and 500,000 K) 
from high resolution instruments
beginning in 1975 have 
been widely interpreted to
be highly dynamic \citep[see the review by ][]{dePontieu+others2021}.  Given
that at least some transition region  emission must originate from electron 
heat conduction down from the 
corona, this has led to 
the expectation that the corona
itself must be dynamic, even in
the quiet Sun.   The impression is that 
the problem of coronal heating is one of 
a highly dynamic system driven by
photospheric motions (subsonic motions with 
time scales of minutes), involving many kinds 
of instabilities and/or flares 
\citep[e.g.][and many references in these works]{Pontin+Priest2022,Raouafi+others2023}.   
Again, any conflict 
between the EUI data analyzed here and transition region
dynamics inferred in multiple
earlier studies is at best, indirect.   In particular we note that
coronal emission lines 
rarely show the signatures in Doppler shifted transition region lines 
of order 100 km~s$^{-1}$,
even though coronal plasma appears to be more tenuous and should
reflect such dynamics.
The transition region also 
has its own basic 
challenges. The entire mass of
the emitting plasmas is so small that modest changes in energy from chromosphere or corona can dominate the dynamics. 

Thirdly, much of the reported evidence for  
dynamics is only 
\textit{indirect}, often
depending on a series of
multiple assumptions
untested or currently untestable.
For example, the review of  \citet{Viall+others2021} includes discussions of
multi-thread, nano-flare driven,  and other empirical constructs in order to account for apparent contradictions of 
data with \textit{ad-hoc} models 
within 
the standard paradigm. 
Should we to accept 
``over-density'' of loops as strong evidence 
for dynamics, and if so, 
on what length- and time-scales?  Without 
delving deeper into 
multiple underlying assumptions in
these earlier studies,
it seems best to draw attention to the 
non-uniqueness of 
 data interpretations
 under the usual paradigm, the limitations of remote sensing, and 
 acknowledge that frequently used  theories (such as fluid mechanics) have no 
 solid prior justification.  Indeed, others have proposed yet-to-be-refuted solutions using assumptions far outside of these models  \citep[e.g.][]{Scudder1992a}). 

Lastly, in a paper entitled
``Dominance of Bursty over Steady Heating of the 4-8 MK Coronal Plasma in a Solar Active Region: Quantification Using Maps of Minimum, Maximum,and Average Brightness''
\citet{Tiwari+others2023} studied the
same kind of 
question with very different 
results.   We might speculate that perhaps the results apply to the far hotter plasmas observed at 9.4 nm in emission from \ion{Fe}{18}
ions formed near 
4-8 MK.  But 
other differences 
in methods (power spectra versus 
reproducible but subjective measures of variability 
on longer time scales), lower
sampling in space and time, make 
meaningful comparisons difficult. It would be interesting to pursue both kinds of data using the same techniques and
data sampling rates.  
 
The arguments presented here contend that the EUI data here are sufficient 
in themselves to prompt new future lines of inquiry, perhaps outside of the usual
set of underlying assumptions. 

\section{Future work}
\label{sec:future}

Beyond these straightforward observations, little quantitative can be said.  Future analysis 
should include a feature identification
algorithm, which
also ``knows'' about 
the amplitudes of variations in time shown
in the figures in this paper.  Machine learning 
techniques agnostic to objects under study would appear to offer a natural
way forward. 
For example, the analysis would benefit
from identifying what
are reasonably identifiable
as transverse oscillations and/or 
field-aligned flows in
the time series, whose only irreversible processes
might include radiation losses, but no 
heating \textit{per se}.
Then, at least, such features
could arguably be ruled out
of the more interesting 
cases where magnetic morphology, and perhaps
topology, is changing.   Such feature classification
would also provide confidence in assigning 
the statistical samples to specific kinds of
structure, hopefully to provide more precise
estimates of properties required by pico-flares
and other mechanisms which remain as yet unrefuted,
in particular near loop footpoints where brightness variations 
are often, but not always, larger than 1\%{}.

One more revision of Parker's
analysis might also be considered. 
Unresolved chromospheric motions, spectroscopic ``turbulence'', exceed
those of the photosphere 
by roughly an order of magnitude 
\citep[e.g.][]{Athay1976,Vernazza+Avrett+Loeser1981, Judge+others2020}.  
But it is unclear yet upon what scale these motions 
occur and if they have cross-field components to generate $B_\perp$
to produce the needed work rate $W$. 
If so, and they exist on large enough scales ($> u\tau$), then again with 
$B\approx100$ G at the top of the chromosphere (this was 
essentially assumed by Parker 
at the coronal base), $u \approx 5$ km~s$^{-1}$, 
 $T\approx 500$ seconds.  
 Then $m=1$, $\tan \theta \approx 0.025$ radians, $B_\perp=2.5$ G, and 
 $\mathcal{E} \approx 10^{23}
$ erg. This consideration points to the need for a
better understanding of the dynamics of plasma 
across the magnetic fields in the upper chromosphere, using DKIST and other large solar telescopes.

The agreement between the analysis from Solar Orbiter EUI data and the numerical
experiments of \citet{Einaudi+others2021} should be explored further.   It will be interesting to see how future 
developments in computing and
observations, confronted 
in the manner presented here, 
might limit the range of coronal heating mechanisms 

Similar datasets from the EUI might also
be examined, especially given the recent 
broad interest in the origins of 
dynamic magnetic fields measured by the Parker Solar 
Probe near 10$R_\odot$. On the basis of
the present paper, these would seem to arise 
only through events related to the emergence
of magnetic flux and accompanying rapid magnetic changes \citep{Raouafi+others2023}.

\section*{Acknowledgments}

The author is grateful to the International
Space Science Institute for a visiting fellowship 
in February 2023 where he was introduced to the 
Solar Orbiter data. He thanks Lucia Kleint and 
the astronomy department at the University of
Bern and the Swiss National Science Foundation 
(grant No. 216870) which made this work 
possible.
 Alin Paraschiv is gratefully acknowledged for comments on the script. 
This
material is based upon work supported by the National Center for
Atmospheric Research, which is a major facility sponsored by the
National Science Foundation under Cooperative Agreement No. 1852977.

\bibliographystyle{aasjournal}
\bibliography{best}
\end{document}